\documentclass[%
 aip,
 jmp,%
 amsmath,amssymb,
 reprint,%
]{revtex4-1}

\usepackage{graphicx}
\usepackage{dcolumn}
\usepackage{bm}
\usepackage[utf8]{inputenc}
\usepackage{graphics,graphicx,dcolumn,bm,fleqn,epic,eepic,float,epsfig}
\usepackage{amssymb,amsmath,multirow,rotate,color,float}
\usepackage{epstopdf,times,color,soul}
\usepackage{csquotes}
\definecolor{red}{rgb}{1,0,0}
\definecolor{green}{rgb}{0,1,0}
\definecolor{blue}{rgb}{0,0,1}

\newcommand{\ie}{\mathrm i}

\begin{document}

\preprint{AIP/123-QED}

\title[]{Bridging between Load-Flow and Kuramoto-like Power Grid Models: A Flexible Approach to Integrating Electrical Storage Units}

\author{Katrin Schmietendorf}
\email{katrin.schmietendorf@uni-oldenburg.de}
 \affiliation{ForWind \& Institut f\"ur Physik, Universit\"at Oldenburg,
          K\"upkersweg 70, 26129 Oldenburg, Germany}
          
\author{O. Kamps}%
\affiliation{ 
Center for Nonlinear Science, Universit\"at M\"unster,
          Correnstraße 2, 48149 M\"unster, Germany
}%

\author{M. Wolff}
\affiliation{Fachbereich Physik, Universit\"at Osnabr\"uck,
          Barbarastraße 7, 49076 Osnabr\"uck, Germany} 

\author{P. G.~Lind}
\affiliation{Fachbereich Physik, Universit\"at Osnabr\"uck,
          Barbarastraße 7, 49076 Osnabr\"uck, Germany}
          
\author{P. Maass}
\affiliation{Fachbereich Physik, Universit\"at Osnabr\"uck,
          Barbarastraße 7, 49076 Osnabr\"uck, Germany}
          
\author{J. Peinke}
\affiliation{ForWind \& Institut f\"ur Physik, Universit\"at Oldenburg,
          K\"upkersweg 70, 26129 Oldenburg, Germany}     

\date{\today}

\begin{abstract}
In future power systems, electrical storage will be the key technology for balancing feed-in fluctuations. 
With increasing share of renewables and reduction of system inertia, the focus of research expands towards short-term grid 
dynamics and collective phenomena. Against this backdrop, Kuramoto-like power grids have been established as a sound mathematical 
modeling framework bridging between the simplified models from nonlinear dynamics and the more detailed models used in electrical engineering.
However, they have a blind spot concerning grid components, which cannot be modeled by oscillator equations, and hence do not allow
to investigate storage-related issues from scratch. 
We remove this shortcoming by bringing together Kuramoto-like and algebraic load-flow 
equations. This is a substantial extension of the current Kuramoto framework with arbitrary grid components.
Based on this concept, we provide a solid starting point for the integration of flexible storage units enabling to address current problems like smart storage control,
optimal siting and rough cost estimations.
For demonstration purpose, we here consider a wind power application with realistic feed-in conditions. 
We show how to implement basic control strategies from electrical engineering, give insights into their 
potential with respect to frequency quality improvement and point out their limitations by maximum capacity and finite-time response.
\end{abstract}

\maketitle

\section{Introduction}
\indent
The transition of the electrical energy system towards sustainability is paralleled by grid decentralization and increasing percentage
of renewables. This development requires novel grid operation and design concepts. 
Electrical storage will be a key component of future energy systems to balance feed-in variations and mitigate 
power quality problems induced  by stochastic 
renewables \cite{albadi2009electricalPowerSystemsResearch,ibrahim2011energyProcedia,ren2017overviewWindPowerIntermittency}.
Therefore, new research issues emerge concerning 
optimal grid embedding and sizing of storage facilities as well as
smart storage control strategies, which are customized to the specific application purpose and feed-in properties.\\
\indent
Wind and solar have characteristic non-Gaussian statistics over a broad range of time
scales from seasonal and diurnal imbalances down to sub-second fluctuations \cite{anvari2016njp,ren2017overviewWindPowerIntermittency}.
Short-term fluctuations on the second and sub-second scale
are a particular challenge for power system operation, since standard load balancing such as primary control\cite{entsoeHandbookPolicy1} does not operate yet on these time scales. 
As a consequence, frequency quality is significantly 
reduced \cite{albadi2009electricalPowerSystemsResearch,ibrahim2011energyProcedia,schmietendorf2017TurbulentRenewableEnergyProductionGridStabilityQuality}.
This problem is exacerbated by a side effect: as conventional power plants are progressively substituted by renewables, system inertia is decreased and
the grid becomes more sensitive to sudden perturbations in terms of feed-in 
fluctuations \cite{entsoeHPoPEIPSl2017,tielens2012gridInertiaFrequencyControlRenewables}.\\
\indent
Against this backdrop, the focus of power grid research shifts towards short-term dynamics and 
multidisciplinary approaches including self-organization and collective phenomena. This requires a profound mathematical modeling framework
mediating between the simple conceptual models from nonlinear dynamics and the detailed models used for case studies in electrical
engineering. Over the past decade, the Kuramoto-like modeling framework has been established as a suitable instrument for this purpose. It is derived from the original Kuramoto model, which describes the phase dynamics of coupled 
oscillators, in particular the phase transition from incoherence to self-organized 
synchronization \cite{kuramoto1975IntSympMathProblems,strogatz2000physica_d}.
Kuramoto-like models have been used to address various issues of power system dynamics and
topology-stability interplay \cite{filatrella2008epjb,menck2014natcomm,motter2013nature,rohden2012prl,rohden2016CascadingFailuresACgrids,rohden2017CuringCriticalLinks,witthaut2016criticalLinksNonlocalRerouting,wolff2018powerGridStabilityHeterogeneityInternalNodes}.
In a previous work \cite{schmietendorf2017TurbulentRenewableEnergyProductionGridStabilityQuality}, it was shown how the turbulent-like
character of wind feed-in, in particular its intermittency, is directly transferred into frequency and voltage fluctuations. This was confirmed by 
real-world frequency measurements \cite{haehne2018footprintTurbulenceGridFreq}. 
Other recent studies on Kuramoto-like grids with stochastic feed-in investigated the propagation of frequency quality
deterioration \cite{auer2017stabilitySynchronyIntermittentFluctTreeLikePowerGrids,zhang2016} and potential routes to system instability 
\cite{schaefer2017EscapeRoutesFluctDrivenNetworks}. However, the current Kuramoto-like framework does not allow to implement grid components, which are not modelled
by oscillator equations. This shortcoming affects the integration of storage units with arbitrary
control strategies from scratch and hence prevents from fundamental investigations of storage-related issues.\\
\indent 
With our study, we fill this gap. The primary target was to integrate a flexible storage 
model, which does not imply any restrictive assumptions on storage features or control strategies 
beforehand.
For this purpose, we introduce a novel approach by bringing together Kuramoto-like 
differential and algebraic load-flow equations, which are a standard tool in power-flow analysis. 
The general idea of embedding grid components by means of load-flow equations has a broader range of application: it can serve as a starting
point to implement arbitrary grid components into Kuramoto-like power grids, e.\,g. power inverters with various types of control or nodes connecting
different grid levels. This broadens the scope of KM-like models significantly. At the same time,
the modeling framework is still a reduced
approach compared to the detailed models used in electrical engineering and yet simple enough to address power grid dynamics 
from the viewpoint of self-organization and collective dynamics, i.\,e. methods beyond the standard engineering practice.
\\
\indent
For demonstration purpose, we consider frequency quality improvement by means of a storage facility with
limited capacity in a simplified power system subjected to realistic wind feed-in. This application example has been identified as
one of the key issues on the road to power systems with high percentage of wind and solar by electrical engineering 
communities \cite{albadi2009electricalPowerSystemsResearch,jabir2017IntermittentSmoothingWindPowerReview,Li2013BESSsmoothingControlWindPV,ibrahim2011energyProcedia,zhao2015reviewStorageWindPowerIntegrationSupport}. 
In order to provide a guide to the implementation of storage units as a starting point for follow-up research, we demonstrate
basic control strategies adopted from electrical engineering and give insights into their potential and limitations with respect to 
different aspects of frequency quality improvement.\\ 
\indent 
The paper is organized as follows: First, we briefly address electrical storage in wind and solar applications and list the
features of real storage units, which should be implementable into the model. 
Then we outline Kuramoto-like power grid modeling and describe how to integrate storage units by means of 
load-flow equations. After that, we specify the simplified power system with realistic wind power input, which we use in this study.
We close the subsection with an explanation of the frequency quality assessment we use, and how this is related to established electrical 
engineering practice. Then we turn to the application example:
We start with a preliminary performance assessment by considering the ability to ensure stationary operation as function of maximum capacity.
Subsequently, we demonstrate how to implement three basic control strategies, namely: \textit{state-of-charge feedback} reinterpreted as storage 
resource management, \textit{droop control} and \textit{ramp-rate control}. We
investigate their potential with respect to frequency quality improvement. 
It shows that these control concepts have different advantages 
according to their underlying main target.
It is pointed out that the ambition in terms of control strength or tolerance range has to be carefully
adjusted to the storage dimension in order to perform optimally. 
Finally, we demonstrate that these strategies are sensitive against finite-time response and confirm that 
short-term frequency quality applications require storage and control systems with rapid response.
We conclude with a summary of the main results
and give an outlook to storage-related problems, which can now be addressed within the context of Kuramoto-like power grid models.

\section{Model and Methods}

\subsection{Electrical storage}\label{subsubsec:storageEquipment}

Electrical energy storage \cite{luo2015overviewStorageTechApplications} denotes the process of converting surplus electrical power into a 
storable form and reserving it, until it is 
converted back when required. It is commonly categorized
by the form of energy stored, but also in terms of their technical features like response time or capacity, or their 
function. Electrical storage is already or considered as a promising candidate for various wind and solar power applications. 
The type of storage follows its function, or to be more precise, the underlying time scale of power variability.
For long-term storage applications like time-shifting, peak-shaving, seasonal storage and mid-term
frequency control, storage types with large energy dimensions are used, which do not necessarily feature fast response,
e.\,g. pumped hydro, hydrogen-based or compressed air storage. Short-term frequency quality
improvement and power output smoothing require rapid response (ranging from few seconds to 
 milliseconds), which is usually paralleled by smaller energy capacity.
Candidates for this application are flywheels, batteries, superconducting magnetic energy storage and (super) capacitors. 
\cite{diazgonzales2012ReviewStorageTechWindPowerApplications,jabir2017IntermittentSmoothingWindPowerReview}\\
\indent
The storage model to be developed has to meet two requirements: 
On the one hand, simplifications are essential in order to fit the model into the Kuramoto-like framework. 
On the other hand, all relevant characteristics of real storage operation
have to be implementable, 
namely\cite{diazgonzales2012ReviewStorageTechWindPowerApplications,jabir2017IntermittentSmoothingWindPowerReview,luo2015overviewStorageTechApplications}:
\begin{itemize}
 \item[\textbullet] \textit{Efficiency}\\
 In practice, the energy conversion processes can not be realized without losses. The efficiency factor $\eta$ gives the ratio of input to
 output energy. It depends on the type of storage.
 \item[\textbullet] \textit{Maximum energy capacity} and \textit{power rating}\\
 These define the main dimensions of the storage facility. The energy capacity is the maximum energy the storage unit is able to deliver and hence
 serves as an upper limit for the amount of energy stored. The power rating corresponds to the maximum instantaneous supply. 
 \item[\textbullet] \textit{Control strategies}\\
 The storage control strategy determines the storage output at time $t$ as a function of one or more feedback variables. 
 It can be used to manage the storage resources or to provide system services like frequency control and power output smoothing. 
 \item[\textbullet] \textit{Response time}\\
 The storage unit has a finite response time effecting a time delay between the feedback signal and its reaction. 
 The response time depends on the type of storage and the underlying control mechanism. 
\end{itemize}

\subsection{Kuramoto-like power grid models}\label{subsubsec:KMlike_models}

Kuramoto-like grid models are based on networks of synchronous machines with producers (generators) and consumers (motors)
converting mechanical power into electrical power and vice versa. Real and reactive power is transferred among these 
nodes via transmission lines. The topology of the underlying network is condensed in the nodal admittance matrix,
$\{Y_{ij}\}_{i,j=1,..,N}$, with $N$ being the number of nodes.
The common assumption of lossless transmission yields $Y_{ij}\approx \ie \mathfrak{Im}(Y_{ij})=\ie B_{ij}$ with susceptance $B_{ij}$.
Each node $i\,\in\mathcal{M}_\mathrm{grid}$ of the grid is associated with
a complex nodal voltage $\bm E_i=E_i\mathrm e^{\ie \delta_i}$ with $E_i$ being the voltage magnitude and $\delta_i$ the phase with respect to 
a reference frame rotating with nominal frequency. (Hence, $\dot\delta_i=\omega_i=0$ means that node $i$ is at nominal 
frequency.)\\
\indent
The coupled frequency-voltage dynamics
of the synchronous machines are given by \cite{machowski2008book,schmietendorf2014epjsti}:
\begin{subequations}
\begin{align}
 m_i\ddot\delta_i=&\,\gamma_i\dot\delta_i+P_i-\sum_{j=1}^N B_{ij}E_iE_j\sin\delta_{ij}, \label{eq:KM1}\\
 \alpha_i\dot E_i=&\,C_i+\beta_i(\bar E_i-E_i)-E_i\hspace{2cm}\phantom{x}\notag\\ 
&+\chi_i\sum_{j=1}^N B_{ij}E_j\cos\delta_{ij}\label{eq:KM2}.
\end{align}
\end{subequations}
The parameters $m_i$ and $\gamma_i$ denote the total inertia and effective damping, $P_i$ is the mechanical power feed-in or consumption.
$P_{ij}=B_{ij}E_iE_j\sin\delta_{ij}$ is the real power transfer between nodes $i$ and $j$, and the interaction term in 
Eq.\,(\ref{eq:KM2}) is related to reactive power flows.
$\alpha_i$, $C_i$, and $\chi$ can be calculated from 
machine parameters. The term $\beta_i(\bar E_i-E_i)$ mimics a proportional voltage controller, which pulls the voltage
towards its nominal value \cite{schmietendorf_toBePublished}.

\subsection{Integration of storage units by means of load-flow equations}\label{subsubsec:implementingStorage}

We now assume a power network consisting of a set of conventional synchronous machines $\mathcal{M}_\mathrm{syn}$ 
(modeled acc. to Eqs.\.(\ref{eq:KM1}) and (\ref{eq:KM2})) and a set of storage units 
with control equipment $\mathcal{M}_\mathrm{SCU}$. 
The storage units are also associated with nodal voltages $\bm E_i=E_i\mathrm e^{\ie \delta_i}$.
However, their dynamics differ from synchronous machines in that they lack inertia and have no inherent physical relationship 
between frequency and electrical power output. The most direct approach, which does not include any restrictive assumptions
on storage features or control, is to calculate
the phase $\delta_i$, $i\in\mathcal{M}_\mathrm{SCU}$, by solving the algebraic load flow equation \cite{kundur1994book,machowski2008book}
\begin{equation}\label{eq:storageModel}
 P_{\mathrm{SCU},i}^\mathrm{out}=\sum\limits_{j=1}^N B_{ij}E_iE_j\sin\delta_{ij},\label{eq:loadflow_p}
\end{equation}
with the nodal voltage $E_i$ assumed to be constant.
$P_{\mathrm{SCU},i}^\mathrm{out}$ is the power being injected into the grid by storage-control unit.\\
\indent
Load-flow analysis is a standard tool in electrical engineering for power-flow calculations on power networks.
It is can be derived from basic physical relationships given by Kirchhoff's and Ohm's laws.
This approach is particularly qualified as a starting point for the investigation of smart control since the power fed into the 
grid $P_{\mathrm{SCU},i}^\mathrm{out}$ can be determined by arbitrary control strategies and the model can flexibly be complemented 
with the other realistic storage characteristics listed above.
The combination with load-flow equations provides a general method for the straight-forward implementation 
of arbitrary grid components into Kuramoto-like networks. This includes, for example, for power inverters and nodes linking
different voltage levels or microgrid-macrogrid connections.

\subsection{Simplified power system with wind feed-in and storage facility}\label{subsec:theSystem}

\begin{figure*}[t!]
\begin{center}
 \includegraphics[scale=0.66]{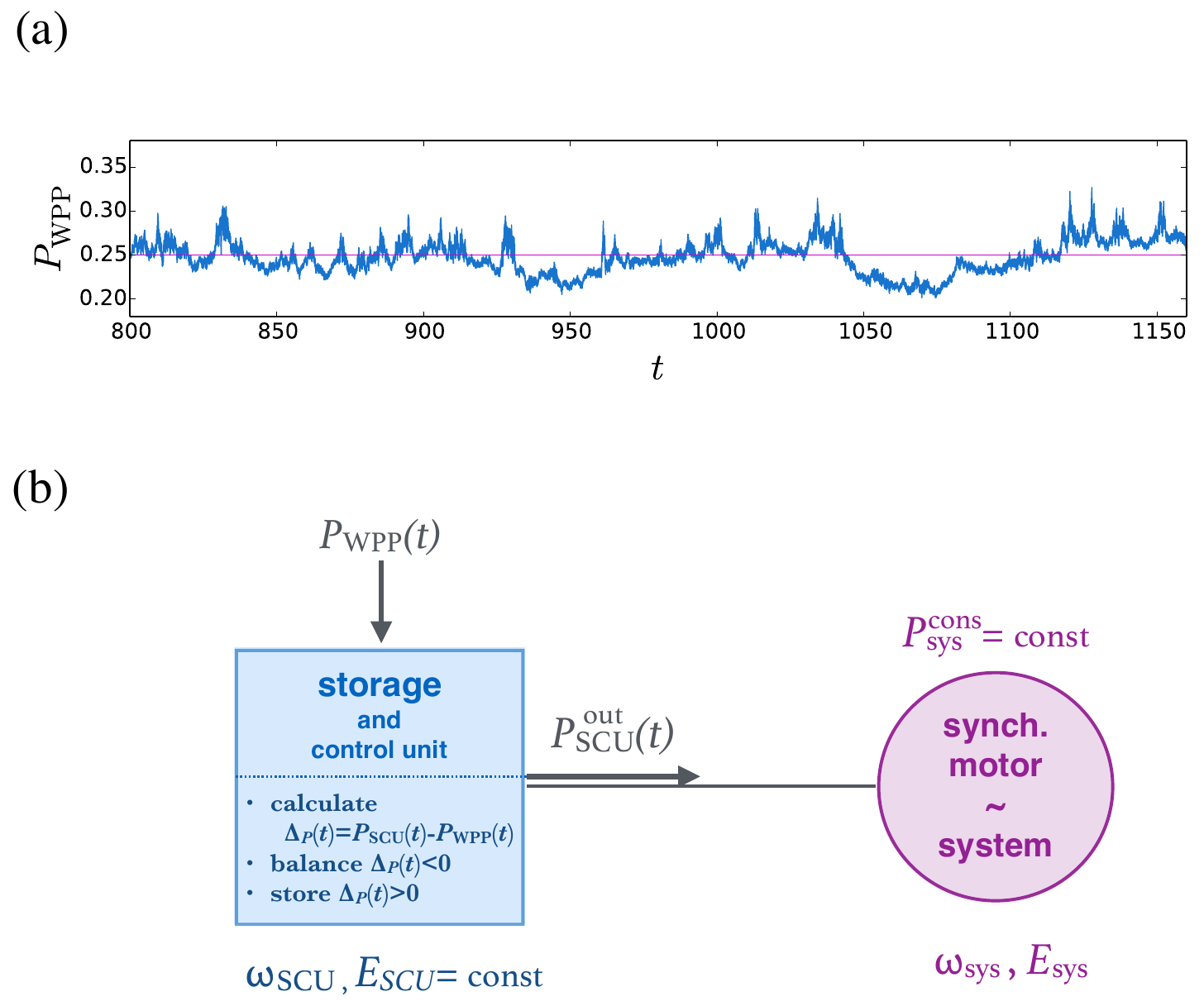}
 \end{center}
 \caption[Simplified model of a power system subjected to wind feed-in with local storage and control.]
 {Simplified model of a power system subjected to wind feed-in with local storage and control.
 (a) Exemplary part of the feed-in time series $P_\mathrm{WPP}(t)=|P_\mathrm{sys}^\mathrm{cons}|+x(t)$ delivered by a wind power plant or park
 (with $|P_\mathrm{sys}^\mathrm{cons}|=0.25$ (magenta line), mean value $\langle x\rangle=0$ and standard deviation 
 $\sigma_x=0.084\cdot|P_\mathrm{sys}^\mathrm{cons}|$). 
 (b) On the basis of the actual wind power feed-in $P_\mathrm{WPP}(t)$, the \underline{s}torage and \underline{c}ontrol
 \underline{u}nit (SCU) first calculates the desired 
 power output value $P_\mathrm{SCU}(t)$ according to the specific control strategy and the mismatch $\Delta_P(t)=P_\mathrm{WPP}(t)-P_\mathrm{SCU}(t)$.
 For $\Delta_P(t)<0$, the mismatch is delivered by the storage, if the filling level $K(t)$ is sufficient. Conversely, surplus power $\Delta_P(t)>0$ 
 can only be stored with $K_\mathrm{max}$ as an upper bound. The power actually fed into the grid by the storage unit is 
 denoted as $P^\mathrm{out}_\mathrm{SCU}(t)\leq P_\mathrm{SCU}(t)$.
 The \enquote{grid node} is modeled as a synchronous machine with parameters 
 (following \cite{schmietendorf2017TurbulentRenewableEnergyProductionGridStabilityQuality}) read: $m=1.0, \gamma=0.2, 
 P_\mathrm{sys}^\mathrm{cons}=-0.25$, $B_{12}=B_{21}=1.0$, $B_{11}=B_{22}=-0.95$, $\alpha=2.0$, $C=0.9101$, $\beta=1.0$, $\chi=0.5$.}
 \label{fig:simpleSystem}
\end{figure*}

In this study, we demonstrate the operation of a storage unit with control equipment by means of a simplified system consisting of a 
generating unit with wind power feed-in and a synchronous machine mimicking the response of the grid 
in terms of frequency $\omega_\mathrm{sys}$ and voltage $E_\mathrm{sys}$ in a coarse-grained view (see Fig.\,\ref{fig:simpleSystem}). 
We consider its application with respect to frequency quality under 
fluctuating wind power feed-in.\\
\indent
The storage unit is assumed to have a finite maximum capacity $K_\mathrm{max}$. 
The actual capacity $K(t)$, or in more casual terms: the \enquote{filling level} at time $t$, corresponds to the 
\textit{state of charge} with respect to batteries.
$K_\mathrm{max}$ serves as an upper bound: if $K(t)=K_\mathrm{max}$, surplus
energy cannot be stored and has to be discarded. On the other hand, the storage unit can only provide balancing power, if $K(t)$ is 
sufficient. For $K(t)=0$, only positive power mismatch can be mitigated, whereas the system is exposed to negative power deficiencies.
\\
\indent
For the sake of simplification, we neglect \textit{efficiency} and limitations due to \textit{power rating} here. This means we assume 
lossless conversion and that the power to be delivered according to the specific control strategy is provided completely if permitted 
by the storage filling $K(t)$. 
The storage unit can be equipped with different control strategies. Three standard strategies adopted from engineering practice
(\textit{state-of-charge dependent resource management}, \textit{droop control}, \textit{ramp rate control}) are specified and investigated
below.
Our intention is to demonstrate the general impact of different basic storage strategies and their limitations due to 
maximum capacity and time-delay on system behaviour rather than to model a detailed situation or derive concrete 
guidelines. The approach can be applied to more concrete situations in the course of follow-up research. 
Complex optimization problems may evolve depending on multiple factors such as operation conditions and technical system requirements,
cost concerns, legal and economic framework etc.

\subsection{Wind power feed-in}\label{subsec:windPowerGeneration}

Due to atmospheric turbulence, wind power has specific turbulent-like characteristics\cite{milan2013prl,anvari2016njp}:  
extreme events, correlations, Kolmogorov
power spectrum, and intermittent increment statistics.
We implement realistic wind feed-in time series taking these basic properties into account 
\begin{equation}
P_\mathrm{WPP}(t)=|P_\mathrm{sys}^\mathrm{cons}|+x(t)\label{eq:feed_in}
\end{equation}
with the constant part $|P_\mathrm{sys}^\mathrm{cons}|$ meeting the consumption of the system. 
The fluctuating time series $x(t)$ is generated as follows \cite{schmietendorf2017TurbulentRenewableEnergyProductionGridStabilityQuality}:
first, a time series $\tilde{x}(t)$
is generated by means of the Langevin-type system of equations 
\begin{eqnarray}
 \dot y&=&-\gamma y+\Gamma(t),\\
 \dot{\tilde{x}}&=&\tilde x\left( g-\frac{\tilde x}{x_0}\right)+\sqrt{D\tilde x^2}y,
\end{eqnarray}
with $\gamma=1.0$, $g=0.5$, $x_0=2.0$, $D=2.0$ and $\delta$-correlated Gaussian white noise $\Gamma$.
Then the corresponding Fourier spectrum is modified so that the final
power spectrum $S(f)=|F(f)|^2$ roughly reproduces real data sets, in particular the Kolmogorov $\frac 53$-decay. Transforming back to real space
yields $x(t)$ (see Fig.\,\ref{fig:simpleSystem}(a) 
for an exemplary part of the feed-in time series $P_\mathrm{WPP}(t)$). Since 
$\langle x\rangle=0$, power balance is given over time: $\langle P_\mathrm{WPP}\rangle=|P_\mathrm{sys}^\mathrm{cons}|$.
Due to the generating process, $P_\mathrm{WWP}$ features a smallest frequency mode. Lower frequencies corresponding to power variations on longer time scales 
are assumed to be handled by other mechanisms like standard load balancing. 
In practice, different time scales can actually be divided up
and assigned to different control mechanisms by low-pass filtering \cite{zhao2015reviewStorageWindPowerIntegrationSupport}.

\subsection{Power quality assessment}\label{subsec:performanceAssessment}

\textit{Power quality} is a wide ranging notion, which includes different aspects of voltage and frequency stability and supply reliability.
In this study, we focus on short-term frequency quality. 
We use different criteria for performance assessment, which in combination give a more detailed picture of frequency 
quality\footnote{Electrical engineering literature indicates \textit{power output smoothing} as 
another major issue with respect to stochastic 
feed-in\cite{jabir2017IntermittentSmoothingWindPowerReview,Li2013BESSsmoothingControlWindPV}. 
This is related to frequency quality are related, as system frequency and power balance are coupled.}.\\
\indent
\textit{Frequency quality} \cite{entsoeHandbookPolicy1,entsoeSupportLoadFreqControl2013,entsoeHPoPEIPSl2017} refers to the systems ability to 
maintain nominal frequency $\omega_\mathrm{sys}^\mathrm{nom}$, or keep the frequency within a pre-defined range 
(the \textit{standard frequency range}) for a large percentage of operation time.
It can be evaluated on different time scales. Short-term frequency quality
is referred to \textit{instantaneous frequency deviations} in electrical engineering. It is commonly evaluated by means of 
the percentage of time the system frequency is outside the standard frequency range \cite{entsoeSupportLoadFreqControl2013}.\\
\indent
Against this practical backdrop, we define
\begin{equation}
 q_{\bar\omega}=
 \begin{cases}
  \int^\infty_{\bar\omega}p(\omega)\mathrm d\omega \qquad\mathrm{for} \,\,\bar\omega>0,\\
     \phantom{\tiny x}\\
  \int^{-\bar\omega}_{-\infty}p(\omega)\mathrm d\omega \qquad\mathrm{for} \,\,\bar\omega<0,
 \end{cases}
\end{equation}
with $\bar\omega$ denoting the bound given by the standard frequency range $\omega_\mathrm{sys}^\mathrm{nom}\pm\bar\omega$ and $p(\omega)$ 
being the
probability distribution of frequency deviations $\omega(t)=\omega_\mathrm{sys}(t)-\omega_\mathrm{sys}^\mathrm{nom}$. 
In real power grids, the
nominal frequency is 50\,Hz or 60\,Hz. Kuramoto-like power grid models are usually transformed into a reference frame rotating with nominal frequency
so that here $\omega_\mathrm{sys}^\mathrm{nom}=0$.
For sufficiently long simulation, $q_{\bar\omega}$ corresponds to the percentage of time the system is expected to operate outside
of the frequency range defined by $\bar\omega$ on average.
With a specified standard frequency range $\bar\omega$, $q_{|{\bar\omega}|}=q_{-\bar\omega}+q_{\bar\omega}$ was introduced
as the \textit{exceedance} measure\cite{auer2017stabilitySynchronyIntermittentFluctTreeLikePowerGrids}.
In this study, we evaluate $q_{\bar\omega}$ as a function
of $\bar\omega$ rather than for one specified $\bar\omega$ for two reasons: firstly, this gives a  more informative picture of system dynamics; secondly, the value of standard frequency range
is not unequivocally defined\footnote{In \cite{entsoeSupportLoadFreqControl2013}, this fact is mentioned with view to the different characteristics of transmission grid areas. 
This applies even more for the heterogeneous operation conditions on the distribution grid level and in islanded microgrids.}.
\\
\indent
Frequency quality not only involves deviations from nominal frequency, but also the time derivative $\mathrm{d}\omega/\mathrm{d}t$, commonly 
referred as the \textit{rate of change of frequency} \cite{entsoeSupportLoadFreqControl2013,entsoeHPoPEIPSl2017}.  
The rate of change of frequency reflects the sensitivity against sudden perturbations and is
inversely proportional to system inertia \cite{tielens2012gridInertiaFrequencyControlRenewables}.
In former times, it was of interest mainly during transient periods after significant imbalances
\cite{entsoeHPoPEIPSl2017}. Nowadays, due to the loss of system inertia and the increasing percentage of stochastic renewables
inducing continuous perturbations, the rate of frequency change becomes
relevant also during \enquote{normal operation}.\\
\indent
The frequency changes can be related to the increments 
$\Delta\omega_\mathrm{sys}=\omega_\mathrm{sys}(t+\Delta t)-\omega_\mathrm{sys}(t)$.
It was shown that intermittency of wind power 
in terms of heavy-tailed probability density functions is directly transferred into frequency fluctuations and significantly 
contribute to frequency quality decrease \cite{schmietendorf2017TurbulentRenewableEnergyProductionGridStabilityQuality}. 
Therefore, we here capture increments statistics not only by their mean value $\mu_{|\Delta\omega|}$ and 
standard deviation $\sigma_{\Delta\omega}$ (as it is standard in electrical engineering) but also their kurtosis 
$\kappa_{\Delta\omega}=\mu_{4,\Delta\omega}/\sigma_{\Delta\omega}^4$
(with $\mu_4$ denoting the fourth moment of the distribution)\footnote{The statistical measures $\mu_{|\Delta\omega|}$, 
$\sigma_{\Delta\omega}$
and $\kappa_{\Delta\omega}$ refer to a given
time lag $\Delta t$, i.\,e. strictly speaking we have $\mu_{|\Delta\omega|}^{\Delta t}$ etc. We drop the upper index, but specify $\Delta t$
in the following analysis}.
The kurtosis serves as a measure for the tailed-ness of the distribution. With $\kappa=3$ being the value of the Gaussian distribution, 
$\kappa>3$ means that there are more extreme events or outliers than in the Gaussian case, and vice versa for $\kappa<3$.  Note that $\kappa$ 
entails information about the shape of the distribution, not about the magnitude of the outliers.\\
\indent
As we will see, the extreme events observed in the increment statistics in this study have two reasons: on the one hand, the system is exposed to the 
feed-in fluctuations $P_\mathrm{WPP}(t)$ during time intervals, in which the storage facility is not able or supposed to fully compensate
power imbalances. As stated above, these fluctuations are known to transfer intermittency into the frequency statistics. 
On the other hand, new extreme events can be induced when the storage steps in. For example, if the system runs out of storage in the course of a longer 
time period with power deficiency or discontinues balancing quite suddenly due to its control specifications,
the frequency may face an instantaneous drop. The following analysis will show that the kurtosis serves as a good indicator for an 
inaccurate adjustment of control strength.

\section{Results}

\subsection{Storage control limited by maximum capacity}\label{subsec:simplestStorage}
We start with a simple storage strategy, which provides a first insight into the performance of the storage facility as a function of its
maximum capacity. 
We assume the storage facility to have a maximum energy capacity $K_\mathrm{max}$. 
The actual power mismatch is $\Delta_P(t)=P_\mathrm{WPP}(t)-P_\mathrm{SCU}(t)$ (see Fig.\,\ref{fig:simpleSystem}).
In this simplified operation mode,
the storage unit is intended to ensure power balance between feed-in and consumption
whenever possible:
\begin{itemize}
 \item[\textbullet] For positive power mismatch $\Delta_P(t)\geq 0$, 
 $P_\mathrm{SCU}^\mathrm{out}(t)=|P_\mathrm{sys}^\mathrm{cons}|$ is fed into the system. The corresponding energy surplus 
$\Delta_K$ is stored with the maximum capacity $K_\mathrm{max}$ being an upper bound.
 \item[\textbullet] For $\Delta_P(t)<0$, the power mismatch is 
 \begin{itemize}
  \item[\textbullet] either fully compensated, i.\,e. $P_\mathrm{SCU}^\mathrm{out}(t)=|P_\mathrm{sys}^\mathrm{cons}|$, if the storage is sufficiently
filled,
  \item[\textbullet] or the rest capacity is used to provide $P_\mathrm{SCU}^\mathrm{out}(t)$ with 
  $P_\mathrm{WPP}(t)< P_\mathrm{SCU}^\mathrm{out}(t)<|P_\mathrm{sys}^\mathrm{cons}|$.
  \item[\textbullet] While $K(t)=0$, no balancing power can be provided.
 \end{itemize}
\end{itemize}

\begin{figure*}[t!]
\begin{center}
 \includegraphics[scale=0.98]{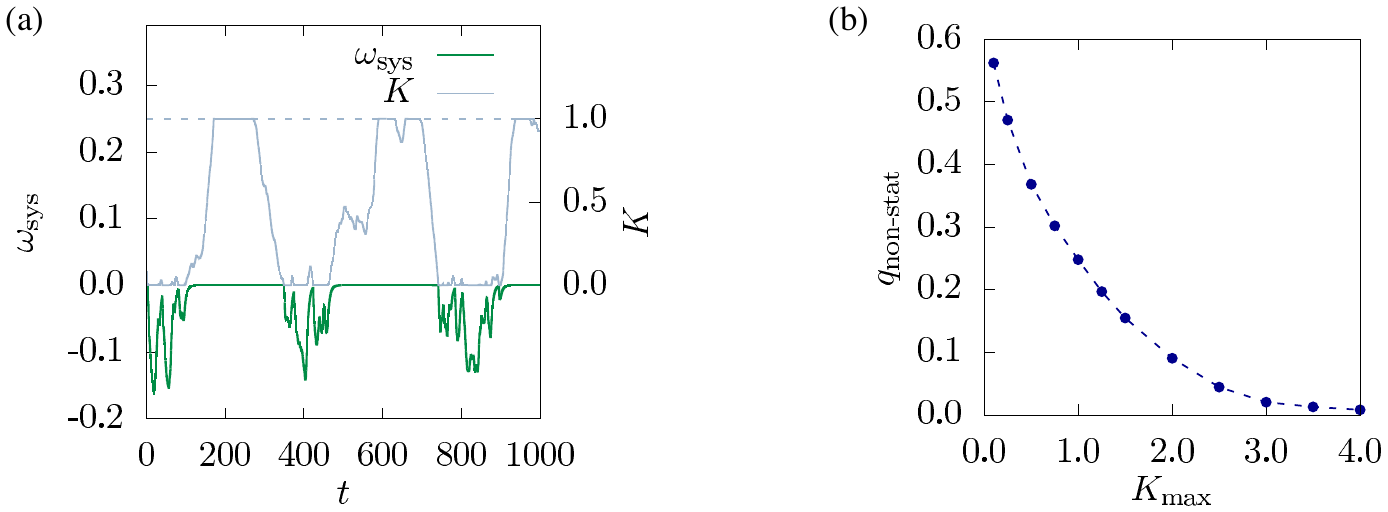}
 \caption[Storage limited by maximum capacity $K_\mathrm{max}$.]
 {Storage limited by maximum capacity $K_\mathrm{max}$. (a) Exemplary time series of system frequency 
 $\omega_\mathrm{sys}(t)$ and storage filling level $K(t)$ for maximum storage capacity 
 $K_\mathrm{max}=1.0$. Empty or insufficient storage filling is paralleled by frequency fluctuations. (b) 
 $q_\mathrm{non-stat}=q_\mathrm{|10^{-3}|}$ gives the percentage of non-stationary operation.}
 \label{fig:maxCapacitiy}
 \end{center}
\end{figure*}

In this mode of operation, the system dynamics alternate between stationary operation and time intervals with $\omega_\mathrm{sys}<0$, 
in which the system either fluctuates in reaction to the stochastic feed-in, or is on its way to return to stationary operation
(see Figure\,\ref{fig:maxCapacitiy}\,(a)).
Figure \ref{fig:maxCapacitiy}\,(b) shows how the percentage of non-stationary operation time 
$q_\mathrm{non-stat}=q_{|\mathrm{10^{-3}}|}$\footnote{We define $|\omega_\mathrm{sys}(t)|<|10^{-3}|$ as stationary 
operation. This is in accordance with real
power grids in the sense that, strictly speaking, these are constantly subjected to disturbances and hence, never in a stationary state
with $\omega_\mathrm{sys}=\omega_\mathrm{sys}^\mathrm{nom}$ and
$\frac{\mathrm{d}}{\mathrm dt}\omega_\mathrm{sys}=0$.} decreases to zero with
maximum storage capacity $K_\mathrm{max}$. The behaviour for the $K\rightarrow\infty$ limit is trivial in qualitative respects, 
as it implies that 
a sufficiently large storage capacity is able to continually balance power differences and guarantee stationary operation. 
It was to be expected as we restricted our analysis to a 
limited time scale, i.\,e. we assumed the feed-in fluctuations to be balanced by other load control mechanisms on longer time scales.\\
\indent
However, our analysis so far was only to serve as a first storage capacity assessment. 
In reality, the equipment of wind and PV plants with large storage capacity is cost expensive. 
In the following, we therefore consider the more
realistic and less trivial situation of a storage facility with \enquote{insufficient capacity}.\\


\subsection{Storage control strategies}\label{sec:controlStrategies}

We now investigate different basic storage control strategies with
regard to their potential to improve frequency quality.
We set the maximum storage capacity $K_\mathrm{max}=2.0$. 
This would allow for stationary operation in about 90 \% of time assuming the simplified
strategy considered above. 
The control strategies refer to accepted methods in engineering practice, and rely on different control
feedback signals:
\begin{itemize}
 \item[\textbullet] the actual storage level $K(t)$, or the \textit{state-of-charge} (here used for the purpose of \textit{storage resource management}),
 \item[\textbullet] system frequency as an indicator for power imbalances (\textit{droop control}), and
 \item[\textbullet] power differences between certain time steps (\textit{ramp rate control}).
\end{itemize}
These methods cover the elementary strategies for power quality improvement and therefore provide a basic structure for 
the development of smart control techniques by refining the conventional strategies or composing hybrid systems.


\subsubsection{Storage resource management}\label{subsec:resourceManagement}

\begin{figure*}[t!]
\begin{center}
 \includegraphics[scale=0.9]{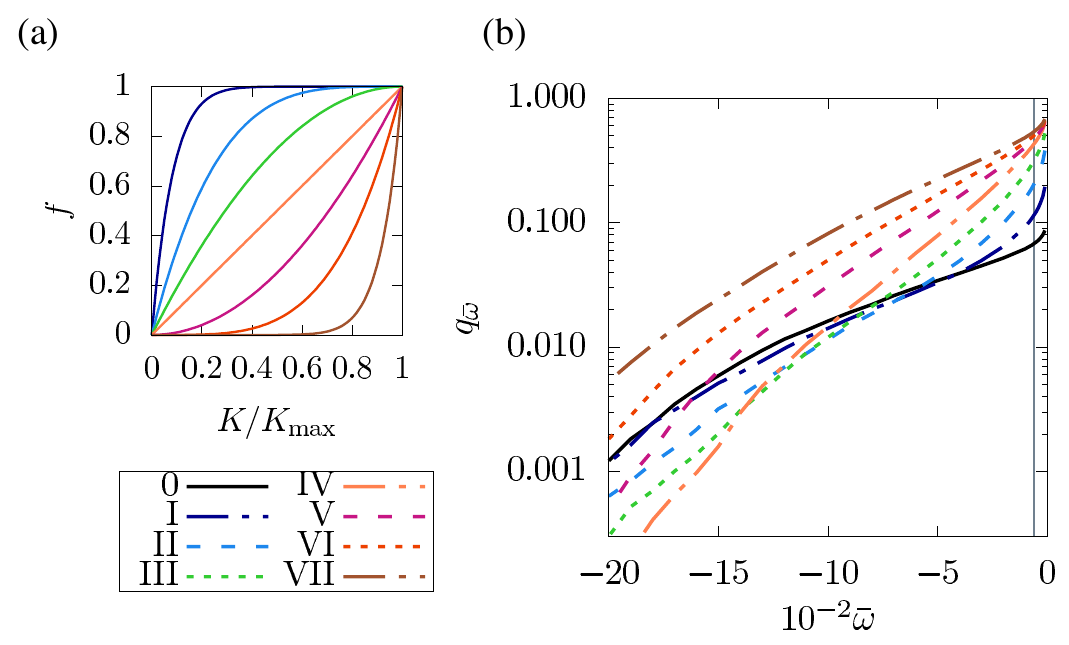}
 \caption[Storage resource management for $K_\mathrm{max}=2.0$.]
 {Storage resource management for $K_\mathrm{max}=2.0$.
 (a) Different realizations of storage resource management $f(K)$ denoted as strategies I-VII. \enquote{0} corresponds to the simplified storage mechanism 
 with $f(K)=1$.
 (b) $q_{\bar\omega}(\bar\omega)$ giving
 the percentage of time the system frequency is outside the $\bar\omega$ boundaries for realizations 0-VII. 
 (As in the case described above, $\omega_\mathrm{sys}(t)\leq 0$ in this mode of operation.)}
 \label{fig:resourceManagement}
 \end{center}
\end{figure*}

In the current technical application, the \textit{state-of-charge} is applied as a feed-back signal in battery storage systems
mainly to guarantee operation within proper state-of-charge
range and to prevent shut-down due to over-charge \cite{Li2013BESSsmoothingControlWindPV}. 
Here, we shift the scope of application to the grid side and reinterpret the basic idea as a form of intelligent storage management.\\
\indent
We complement the simple storage strategy presented above and make the balance power at time $t$ dependent
on the current capacity $K(t)$.
To be specific, the power to be delivered by the storage is the power deficiency $\Delta_P$
multiplied by a factor $f=f(K(t))\in[0,1]$ with $f(0)=0$ and $f(K_\mathrm{max})=1$.
Here, we consider the different realizations for $f(K)$ depicted in Fig.\,\ref{fig:resourceManagement}\,(a):
$f(K)=-(1-K/K_\mathrm{max})^n+1$ for $n=12,\,4,\,2$ (denoted as scenarios I,\,II and III) and $f(K)=(K/K_\mathrm{max})^n$ for 
$n=1,\,2,\,4,\,12$ (scenarios IV-VII).\\
\indent
Fig.\,\ref{fig:resourceManagement}\,(b) shows $q_{\bar\omega}(\bar\omega)$ for the different realization of storage resource management 
in comparison to the simplified storage strategy described in the previous subsection. 
With proper a choice of $f(K)$, the proposed storage resource management can in fact serve to prevent large frequency deviations. 
Of course, a higher percentage of small deviations has to be tolerated in exchange.
The best option can only be chosen in knowledge of the operational circumstances and the specific guidelines for $\bar\omega$. 
Furthermore, storage resource
management can be applied as part of a multi-pronged control strategy in combination with other storage control mechanisms.

\subsubsection{Droop control}\label{subsec:droopControl}

\begin{figure*}[t!]
\begin{center}
\includegraphics[scale=0.86]{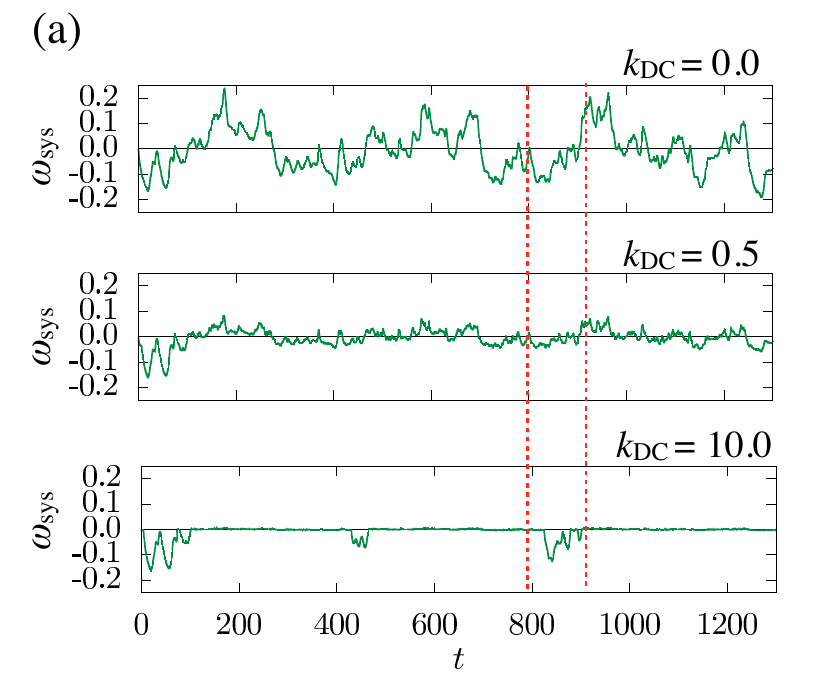}\\
\includegraphics[scale=0.76]{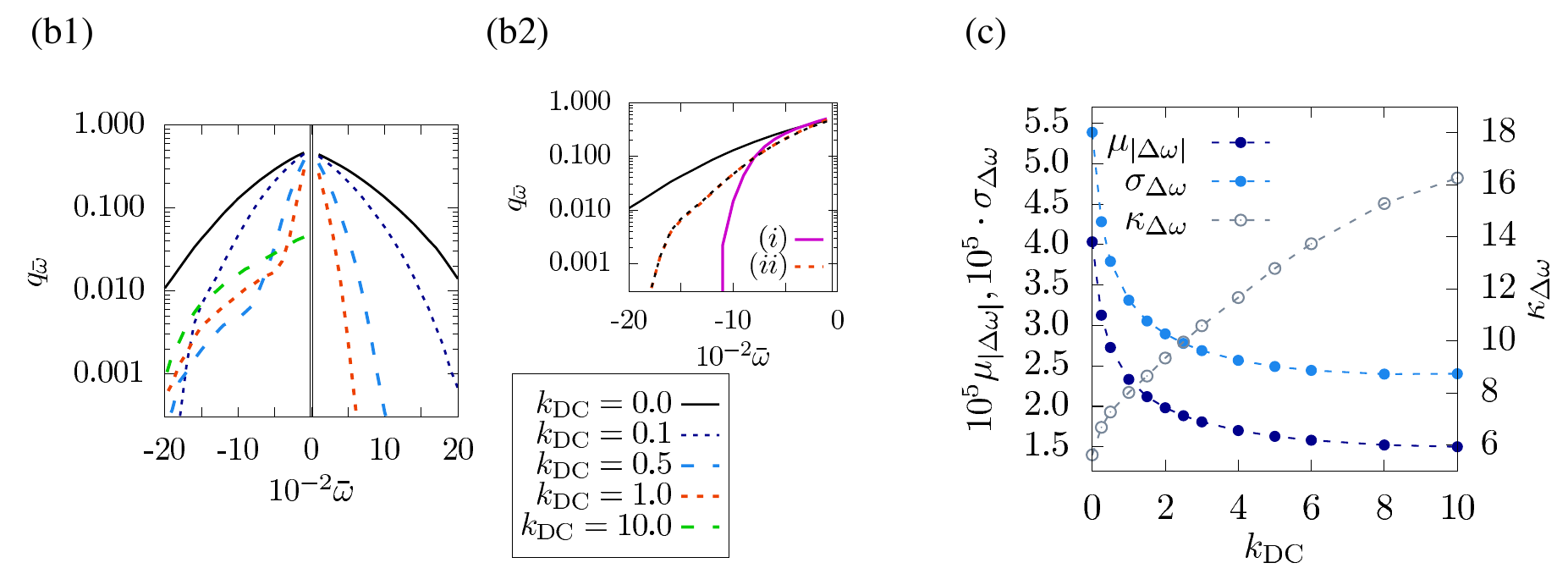}
 \caption[Droop control with maximum storage capacity $K_\mathrm{max}=2.0$.]
 {Droop control with maximum storage capacity $K_\mathrm{max}=2.0$. (a) System frequency response to the same feed-in time series 
 with standard droop control acc. to Eq.\,(\ref{eq:droopControl})
 for different control strengths $k_\mathrm{DC}=0.0$ (no control), $k_\mathrm{DC}=0.5$, and $k_\mathrm{DC}=10.0$. The time interval indicated by
 the red dotted lines illustrates the drawback of too ambitious control strength: the frequency dip is prevented for $k_\mathrm{DC}=0.5$,
 but no longer for $k_\mathrm{DC}=10.0$.
 (b1) $q_{\bar\omega}(\bar\omega)$ for standard droop control with different control strengths $k_{DC}$. For 
 $k_\mathrm{DC}=10.0$ the curve for positive deviations is not displayed due to its rapid decay ($q_{0.01}$ has already dropped
 to $\mathcal{O}(10^{-5})$).
 (b2)  $q_{\bar\omega}(\bar\omega)$ for the alternative non-symmetric droop control
 strategies acc. to Eq.\,(\ref{eq:droopControl_mod}): (i) $n=4$, $k^\mathrm{DC}_2=10.0$, $k^\mathrm{DC}_2=200.0$, and (ii) 
 $n=1$, $k^\mathrm{DC}_2=10.0$, $k^\mathrm{DC}_2=0.1$. For negative $\bar\omega$, (ii) resembles the standard droop control case.
 (c) Increment statistics during non-stationary operation (according to the definition given above). For increasing $k_\mathrm{DC}$ in standard droop control, 
 the mean value $\mu_{|\Delta\omega|}$ and standard deviation $\sigma_\omega$
 decrease, while the non-Gaussianity of the distribution in terms of the kurtosis $\kappa_{\Delta\omega}$ grows.}
 \label{fig:droopControl}
 \end{center}
\end{figure*}

\textit{Droop control} is based on the relationship between power imbalances and system frequency:
a positive power mismatch is paralleled by frequency increase, 
whereas negative mismatch leads to frequency decrease. 
This fact can also be observed in Kuramoto-like grids.\\
\indent 
Droop control has a broad range of application, which
includes frequency control services provided by wind power plants \cite{zhao2015reviewStorageWindPowerIntegrationSupport}.
The standard practice is to use a linear droop control mechanism, whose slope is given by the control strength $k_\mathrm{DC}$. 
The balancing droop power is 
\begin{equation}\label{eq:droopControl}
 P_\mathrm{droop}(t)=k_\mathrm{DC}(\omega_{\mathrm{sys}}^\mathrm{nom}-\omega_\mathrm{sys}(t))=-k_\mathrm{DC}\omega(t),
\end{equation}
as $\omega_{\mathrm{sys}}^\mathrm{nom}=0$ here.
If $\omega_\mathrm{sys}(t)<0$, this is interpreted as an indicator of negative power balance, and consequently more
power is injected into the system in order to keep system frequency close to its nominal value. 
For $\omega_\mathrm{sys}(t)>0$, power feed-in is reduced accordingly.\\
\indent
This adjustment of power input to the actual system frequency obviously requires storage capacity in the background.
A specific type of inverter with linear droop control was shown to behave analogue to a synchronous machine \cite{schiffer2013SyncDroopControlledMicroGrids}, and
was already implemented into Kuramoto-like grids with stochastic feed-in \cite{auer2017stabilitySynchronyIntermittentFluctTreeLikePowerGrids}. 
Our approach here is different 
in the sense that we explicitely
take into account the limits of the installed background storage capacity $K_\mathrm{max}$ but
do not assume any further specifications on the grid feed-in process. 
As explained above, in case of $\omega_\mathrm{sys}(t)<0$, the balancing power $P_\mathrm{droop}>0$ can only be provided 
to the extent that the storage level $K(t)$ is 
sufficient, and for $\omega_\mathrm{sys}(t)>0$, surplus power can only be stored with $K_\mathrm{max}$ as an upper bound.\\
\indent
We first investigate system performance under standard droop control according to Eq.\,(\ref{eq:droopControl}) for fixed maximum storage capacity 
$K_\mathrm{max}=2.0$ and varying control strength $k_\mathrm{DC}$. 
Fig.\,\ref{fig:droopControl}(a) shows system frequency in response to the same power feed-in $P_\mathrm{WPP}(t)$ for different $k_\mathrm{DC}$.
With increasing control strength, the positive frequency
deviations are more and more eliminated, as it is always possible to feed in less power than available.
In contrast, balancing negative frequency deviations requires sufficient storage level. 
This asymmetry can also be seen in
Fig.\,\ref{fig:droopControl}(b1), which reveals the dilemma of standard droop control under limited storage capacity: On the one hand, for 
sufficiently large control strength $k_\mathrm{DC}$, the positive frequency deviations can be more or less eliminated; but in this case, the storage unit 
runs out of capacity quickly at the beginning of longer timer periods with negative power mismatch. 
Fig.\,\ref{fig:droopControl}(a) highlights an concrete example of a frequency dip not being prevented due to overambitious control strength.
On the other hand, for small control strength, the storage facility performs better in the sense that it 
reduces the probability of large negative frequency deviations. 
But at the same time, the droop control mechanism remains sub-optimal with respect to positive frequency deviations.\\
\indent
To overcome this problem, we propose a non-symmetric droop control strategy, which treats positive an negative frequency 
deviations differently:
\begin{equation}\label{eq:droopControl_mod}
P_\mathrm{droop}(t)=\begin{cases}
    -k^\mathrm{DC}_1\omega_{\mathrm{sys}}(t)\qquad \forall\,\omega_{\mathrm{sys}}\geq0,\\   
    k^\mathrm{DC}_2(\omega_{\mathrm{sys}}(t))^n\qquad \forall\,\omega_{\mathrm{sys}}<0.
   \end{cases}
\end{equation}
We choose large control strength $k^\mathrm{DC}_1$ in order to counteract the positive frequency deviations 
and tested two control schemes for the negative frequency range: 
(i) quartic droop control\footnote{Nonlinear droop control is technically feasible \cite{nonLinDroopControl2018ieee}.} 
; and (ii) linear droop control with small control strength $k^\mathrm{DC}_2$.
Fig.\,\ref{fig:droopControl}(b2) shows that these alternative strategies combine the best of both small and large 
control strength in standard droop control: they effectively mitigate positive deviations and prevent
large frequency dips. 
A nonlinear control-term, inter alia, gives the opportunity to focus the onset of control to a specific frequency bound. 
For example, the drop of $q_{\bar\omega}(\bar\omega)$ indicates that the quartic control term actually starts acting around 
$\omega_{\mathrm{sys}}\approx 0.05$.\\
\indent
With view to the frequency increments statistics (see Fig.\,\ref{fig:droopControl}(c)), 
the mean value $\mu_{|\Delta\omega|}$ and standard deviation $\sigma_{\Delta\omega}$ decrease with increasing control strength, finally
converging to a minimum value. However, the non-Gaussianity in terms of kurtosis $\kappa_{\Delta\omega}$ grows, even when $\mu_{|\Delta\omega|}$ and
$\sigma_{\Delta\omega}$ have nearly approached their minima and barely change\footnote{The corresponding values of the statistical measures for the alternative strategies (i) and (ii) 
are: (i) $\mu_{|\Delta\omega|}=2.25\cdot 10^{-5}$, $\sigma_{\Delta\omega}=3.21\cdot 10^{-5}$, $\kappa_{\Delta\omega}=9.00$, and 
 (ii) $\mu_{|\Delta\omega|}=2.26\cdot 10^{-5}$, $\sigma_{\Delta\omega}=3.16\cdot 10^{-5}$, $\kappa_{\Delta\omega}=8.50$.}.
This is an indicator that the control strength $k_\mathrm{DC}$ is getting too ambitious and the storage facility runs out of capacity
more frequently. It therefore becomes evident that increment statistics are an essential part of a comprehensive
picture of frequency quality.

\subsubsection{Ramp rate control}\label{subsec:rampRateControl}

\begin{figure*}[t!]
\begin{center}
 \includegraphics[scale=0.92]{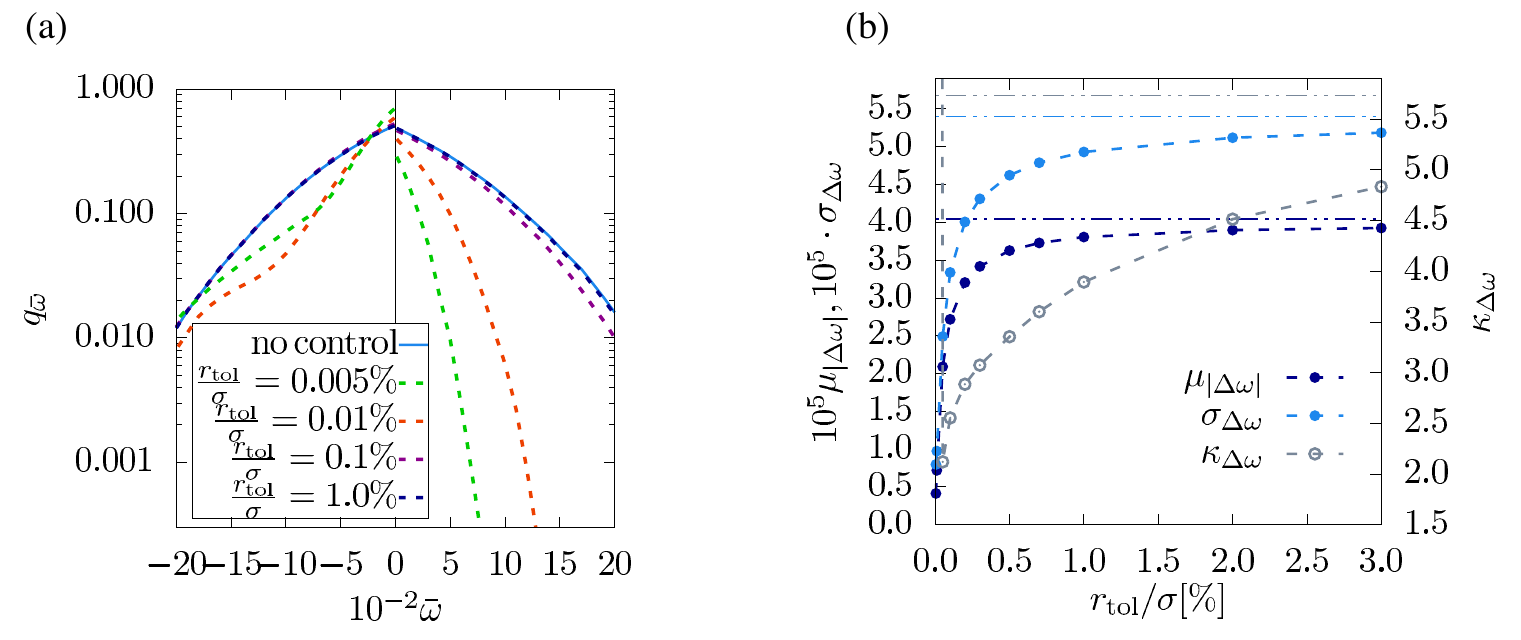}
 \caption[Ramp rate control with $K_\mathrm{max}=2.0$.]
 {Ramp rate control with $K_\mathrm{max}=2.0$. (a) $q_{\bar\omega}(\bar\omega)$ for different tolerance ramp rates
 $r_{tol}$ given in percent of the standard deviation of the wind power feed-in $\sigma$. 
 (b) Increment statistics during non-stationary operation: with decreasing ramp tolerance $r_\mathrm{tol}$, frequency increments are mitigated in terms of their
 mean value $\mu_{\Delta\omega}$, standard deviation and kurtosis $\kappa_{\Delta\omega}$.  For large tolerance
 ramps, the influence of control diminishes and $\mu_{\Delta\omega}$, $\sigma_{\Delta\omega}$ and $\kappa_{\Delta\omega}$ approach their values 
 of the no-control case (horizontal lines).}
 \label{fig:rampRateControl}
 \end{center}
\end{figure*}

A \textit{power ramp} is defined as a normalized power change or power increment:
\begin{equation}
 \Delta P_\mathrm{ramp}(t)=\frac{P_\mathrm{in}(t)-P_\mathrm{ref}}{P_\mathrm{norm}}
\end{equation}
with input power $P_\mathrm{in}(t)$ and reference power $P_\mathrm{ref}$. 
\textit{Ramp rate control} \cite{marcos2014storageRequirementsPVrampRate,schnabel2016storageRequirementsPVrampRateNorthernEurope} 
aims at keeping power ramps within specified tolerance bounds: 
\begin{equation}\label{eq:rampCondition}
 |\Delta P_\mathrm{ramp}|\leq r_\mathrm{tol}.
\end{equation}
It is utilized in wind and solar power applications. In the latter case, power ramps play a even major role due to passing clouds.\\
\indent
The basic idea opens up numerous opportunities for concrete realization depending of the choice of $P_\mathrm{ref}$. For example, it can 
be given by prior values $P(t-\Delta t)$ defined by a sampling time
$\Delta t$ or be calculated as a function of the actual demand.
We here demonstrate a version of ramp rate control, which mainly targets on short-term ramps:
First, we set $P_\mathrm{ref}=P_\mathrm{SCU}^\mathrm{out}(t-\Delta t)$ with $\Delta t=0.005$.
As long as the ramp condition Eq.\,(\ref{eq:rampCondition}) is satisfied, no balancing is necessary and
$P_\mathrm{SCU}^\mathrm{out}(t)=P_\mathrm{WPP}(t)$. If the condition is violated, the storage facility steps in: For $\Delta P_\mathrm{ramp}>0$ 
(upward ramps),
$P_\mathrm{SCU}^\mathrm{out}$ is decreased so that $|\Delta P_\mathrm{ramp}|=r_\mathrm{tol}$ and surplus power is stored.
For $\Delta P_\mathrm{ramp}<0$ (downward ramps), the storage is supposed provide balance power in order to fulfill 
$|\Delta P_\mathrm{ramp}|=r_\mathrm{tol}$.
Again, the storage of surplus power is limited by the maximum capacity $K_\mathrm{max}$ and balancing power can only be delivered if
the actual storage level $K(t)$ is sufficient.\\
\indent
The performance of ramp rate control is usually assessed with respect to the power feed-in statistics. Here, 
we consider frequency statistics instead, for two reasons: first, this is the scope of our study and consistent with the previous analysis.
Secondly, we investigate a consequential phenomenon, as power fluctuations are directly transferred into frequency variations.
Fig.\,\ref{fig:rampRateControl} (a) and (b) show how the likelihood of tolerance bound violations in terms of $q_{\bar\omega}(\bar\omega)$ and
the frequency increment statistics evolve as functions of the tolerance ramp rate $r_\mathrm{tol}$. 
It shows that the ramp rate control strategy fulfills its 
main purpose with view to the increment statistics: by suppressing power ramps, frequency increments can be mitigated significantly. 
In parallel, the percentage of operation time
beyond certain tolerance bounds can be decreased.\\
\indent 
Again, the ambition of control, here in terms of $r_\mathrm{tol}$, has to be chosen carefully. 
On the one hand, if $r_\mathrm{tol}$
is too large, the control does not achieve its potential. It has no influence on $q_{\bar\omega}(\bar\omega)$ and barely improves the 
increment statistics. On the other hand, if $r_{tol}$ is too 
small, the storage tends to run out of capacity. This is indicated by a steep rise of $\kappa_{\Delta\omega}$ and
increasing likelihood of large negative frequency deviations.
Compounding the problem in this specific version of \textit{ramp-rate control} is the fact that 
if the power input $P_\mathrm{SCU}^\mathrm{out}$ drops to a low $P_\mathrm{WPP}$ during a feed-in deficit period with empty storage,
this value serves as the new reference $P_\mathrm{ref}$. In the following, the input power and system frequency can return to their nominal values
only slowly due to the tight tolerance range, even if storage capacity is available. As explained before, running out of 
storage is paralleled by sudden frequency drops, which is indicated by the drastic increase of the non-Gaussianity of the increment distribution.\\
\indent
Note that the dissymmetry between positive and negative frequency deviations 
(which can be seen in Fig.\,\ref{fig:rampRateControl}\,(a)) is not completely analogue to the droop control case. First, ramp rate
control responds to power input fluctuations (which cause of frequency fluctuations) and not to the deviation from nominal frequency 
directly. 
Secondly, upward ramps
can always be balanced, irrespective whether the actual system frequency is below or above its nominal value. In contrast, 
balancing downward ramps requires sufficient storage. This particularly affects 
power deficit periods accompanied by $\omega_\mathrm{sys}<0$, during which the storage is depleted.\\
\indent 
Comparing droop control and the applied version of ramp rate control, the latter has the advantage to be
able to mitigate frequency increments to a certain extent without being paralleled by increasing non-Gaussianity. For example, droop control
with $k_\mathrm{DC}=6.0$ and ramp control with $r_{tol}=0.03\sigma\%$ both reduce the mean value  to 
$\mu_{\Delta\omega}\approx 1.6\cdot 10^{-5}$. At 
the same time, the statistics for the droop control case contain considerably more extreme events ($\kappa_{\Delta\omega}=13.7$) than the
system with ramp rate mechanism ($\kappa_{\Delta\omega}=1.8$).
On the other hand, the ramp rate control does not take into account the absolute deviation from nominal frequency, and hence is 
not designed to prevent large frequency excursion as efficiently as droop control.

\


\subsection{Finite response time}

\begin{figure*}[t!]
\begin{center}
 \includegraphics[scale=0.78]{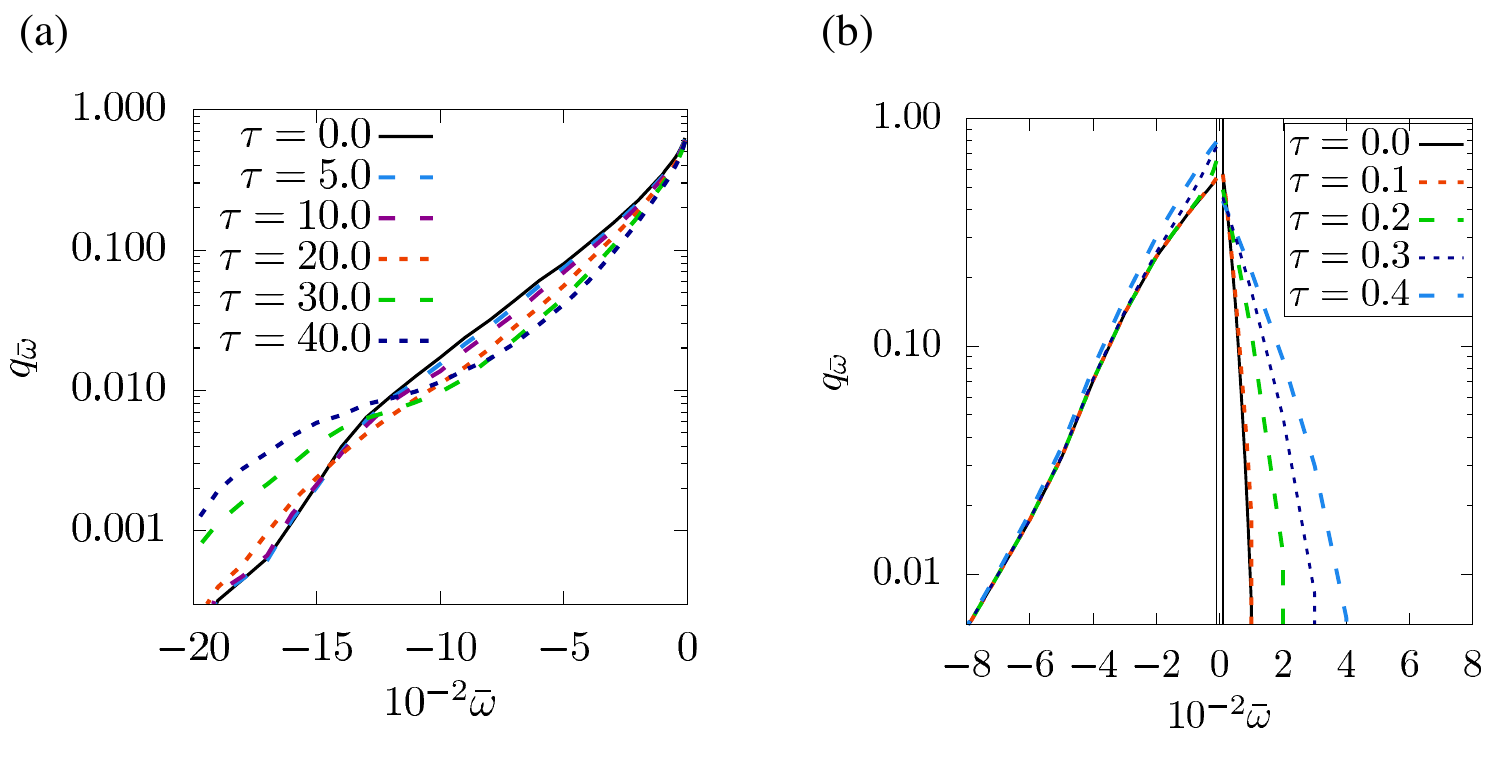}\\
  \includegraphics[scale=0.78]{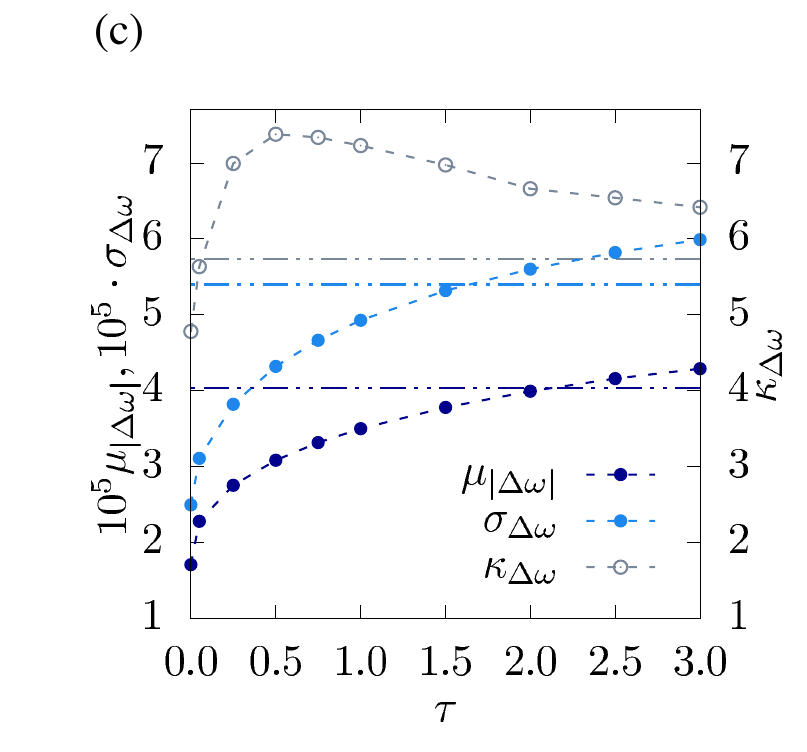}
 \caption[How finite response time undermines frequency quality improvement.]
 {How finite response time undermines frequency quality improvement.
 (a) Linear storage resource management: $q_{\bar\omega}(\bar\omega)$ for different 
 response times $\tau$. The curves for $\tau<5.0$ almost resemble the instantaneous-response ($\tau=0$) case. (b) Droop control for the 
 asymmetric control strategy (ii) . Again, $q_{\bar\omega}(\bar\omega)$ is shown 
 for different response times. For $\tau<0.1$, the impact of finite-time response on $q_{\bar\omega}(\bar\omega)$ is negligible. (c) 
 Increment statistics for ramp rate control as a function of $\tau$. The horizontal lines indicate the values for the no-control case.}
 \label{fig:timeDelay}
 \end{center}
\end{figure*}

Real control equipment does not react instantaneously but in response 
to the feedback signal at time $t-\tau$. In the following, we investigate and compare the sensitivity of the three control strategies 
introduced in the previous section. 
We implemented finite time response as follows\footnote{Alternatively, the time delay could be modelled by
an ordinary differential equation with an appropriate time constant.}:
\begin{itemize}
 \item[\textbullet] In case of \textit{storage resource management}, the balance power to be delivered by the storage facility is 
 $\Delta_P\cdot f(K(t-\tau))$\footnote{We here assumed finite-time response solely for the filling-level feedback, while the calculation 
 of $\Delta_P$ and hence the decision
whether the storage facility has to step in, happens instantaneously in response to the actual mismatch $\Delta_P(t)$. If this was not 
the case (this process would perhaps be associated with another
time delay $\tau^\prime$), the situation would of course be exacerbated and also positive frequency deviations could occur.}.
We picked the linear storage resource management scenario VI as 
 example.
  \item[\textbullet] For \textit{droop control}, we instance the asymmetric control strategy (ii) 
  with $k^\mathrm{DC}_1=10.0$ and $k^\mathrm{DC}_2=0.5$. 
  The balancing power $P_\mathrm{droop}(t)$ is calculated on the basis of 
  $\omega_\mathrm{sys}(t-\tau)$. 
  \item[\textbullet] The \textit{ramp-rate control} realization we presented above is very sensitive due to the
  short sampling rate. In view of finite time response, we consider another variant of ramp rate control and define the reference power  
  $P_\mathrm{ref}=|P_\mathrm{sys}^\mathrm{cons}|$ $\forall t$ and the tolerance range $r_\mathrm{tol}=0.5\sigma$ here. 
  The balancing power at time $t$ is calculated as the response to $\Delta P_\mathrm{ramp}(t-\tau)$.
\end{itemize}
Fig.\,\ref{fig:timeDelay} shows how time delay limits the possibilities for frequency quality improvement 
with focus on the main target of each control strategy.\\
\indent
Storage resource management was introduced in order to prevent overspending and save capacity to mitigate large frequency deviations.
From Figure\,\ref{fig:timeDelay}\,(a), one can see that finite response time has negligible impact up to $\tau=10.0$.
Then the deviations from the instantaneous-response case become more and more apparent. In particular,
the control strategy increasingly misses its main objective as the probability of large deviations from nominal frequency grows.\\
\indent
Droop control is intended to mitigate deviations from nominal frequency.
Fig.\,\ref{fig:timeDelay}\,(b) shows that in this respect the system is able to handle a finite response time up to $\tau=0.1$ quite well.
Then $q_{\bar\omega}$ starts to increase for small $\bar\omega$ as
the feedback delay causes trouble when the system fluctuates close around
nominal frequency. As the control switches between positive and negative balancing too late, oscillations around nominal frequency are
induced.
A nonlinear droop scheme could mitigate these oscillations as it interfers less for small deviations.\\
\indent
Ramp rate control was shown to be a promising candidate for frequency quality improvement with respect to increment statistics.
The version of ramp-rate control considered here is very sensitive towards time delay. 
Fig.\,\ref{fig:timeDelay}\,(c) shows that the introduction of finite response time leads to a reduction of frequency 
quality as $\mu_{|\Delta\omega|}$, $\sigma_{\Delta\omega}$ and $\kappa_{\Delta\omega}$ immediately increase, 
even beyond the no-control case.\\

\indent
These results indicate that short-term frequency quality applications require rapid response of the underlying control 
mechanism (on sub-second scale\footnote{As we use a 
greatly simplified power system in dimensionless units in order to point out general relationships, we are careful with 
specifying concrete values. However, at this point, we want to give a rough idea about the time scale. For 
$\mathcal{O}(m)\sim10^4\,\mathrm{kg\,m}^2$,
$\omega^\mathrm{nom}_\mathrm{sys}=2\pi\cdot 50\,$Hz, $\mathcal{O}(BE_\mathrm{sys}E_\mathrm{SCU})\sim 1\,\mathrm{GW}$ and 
$\mathcal{O}(1/\gamma)\sim 0.1\,\mathrm{s}-1\,\mathrm{s}$
(cf.\,\cite{menck2014natcomm}), $t=\mathcal{O}(t)\sim 0.1\,\mathrm{s}-1\,\mathrm{s}$ and hence the response time of the storage facility has to be
in the sub-second range. This coarse evaluation is in accordance with the technical features of electrical 
storage with the fastest response times being in the range of milliseconds \cite{zhao2015reviewStorageWindPowerIntegrationSupport}.}). 
As this is usually cost-expensive, it may be advisable to use hybrid systems and treat the 
high-frequency and lower frequency fluctuations separately with different storage and control systems.


\section{Discussion and Outlook}
We extended the current Kuramoto-like modeling framework with flexible storage units. 
With that, the scope of Kuramoto-like models opens up to one of the most important research topics in power grid engineering.
On the way to this goal, we brought together
Kuramoto-like equations and load-flow analysis. This is a substantial extension, which 
can serve as a starting point for the straight-forward 
implementation of arbitrary grid components. \\
\indent
For demonstration purposes, we considered short-term frequency quality 
improvement by means of storage facility with maximum capacity in a power system subjected to realistic wind feed-in.
Motivated by recent findings, we assessed system performance not only with respect to frequency range violations, but also 
took into account frequency increment statistics.\\
\indent 
We demonstrated how to implement three basic control methods, which cover the elementary strategies 
for power quality improvement in engineering practice.
First, we adopted \textit{state-of-charge feedback control} and reinterpreted it as a form of storage resource management. 
It has been proven that this concept can actually serve to save capacity in order to prevent large frequency deviations. 
Secondly, it was shown that \textit{droop control} can improve frequency quality not only with view to deviations 
from nominal frequency but also with respect to frequency increment statistics. We pointed out that, particularly in case of limited capacity,
it is 
favorable to handle positive and negative frequency deviations with different droop schemes and consider non-linear mechanisms.
Thirdly, we implemented a version of \textit{ramp rate control}. Originally designed for power-output-smoothing 
applications, we demonstrated that this strategy entails frequency quality improvement.\\
\indent
For both droop and ramp-rate control, it became apparent that
the corresponding control strength or ramp tolerance range may not be too ambitious and have 
to be carefully proportioned to the dimensions of the storage facility. Furthermore, it was shown that the finite response time of the control
mechanism limits the potential of the storage facility. Short-term frequency quality applications in particular require a rapid response.
\\
\indent
With this study, we created a sound starting point for follow-up research on various aspects of storage implementation
from the viewpoint of self-organized synchronization and collective phenomena. 
This includes stability-topology issues like 
\textit{optimal siting} of storage units as well as comparative studies on global vs. local storage location or \textit{optimal sizing} and
rough cost-benefit assessment. 
Another current topic is the development and refinement of smart control strategies, 
which are customized to the realistic features of wind and solar power and, at the same time, take into account the impact of
collective network dynamics. This study has already shown that 
the presented basic control strategies have different advantages and disadvantages. Against this backdrop, and with view to the impact of
finite response times, systems with combined 
control techniques are conceivable solutions and novel smart control strategies should be developed.



\section*{Acknowledgements}
Financial support from the Deutsche Forschungsgemeinschaft (PE 478/16-1 and MA 1636/9-1) is gratefully acknowledged.

\nocite{*}
\bibliography{KM_storage_arxiv}

\end{document}